\documentclass[seceq]{ptptex}

\usepackage[dvips]{graphicx,color}
\usepackage{epsfig}
\usepackage{bm} 
\usepackage{amsmath}
\usepackage{amssymb}
\usepackage{booktabs}

\notypesetlogo

\newcommand{\ev}[1]{\left\langle{}{#1}\right\rangle{}}

\renewcommand{\vec}[1]{\bm{#1}}
\newcommand{\nn}{\nonumber}

\newcommand{\del}[2]{\frac{\partial #1}{\partial #2}}

\newcommand{\tsin}{\tilde{s}}
\newcommand{\tcos}{\tilde{c}}
\newcommand{\comb}[2]{{_{#1}C_{#2}}}
\allowdisplaybreaks[4]

\begin{document}

\title{Laser Irradiated Enhancement of the Atomic Electron Capture Rate in search of New Physics}

\author{Takaaki Nomura, Joe Sato and Takashi Shimomura}
\inst{Department of Physics, Saitama University, 
	Shimo-okubo, Sakura-ku, Saitama, 338-8570, Japan}

\date{\today}

\abst{
Electron capture processes are important in the search for new physics. In this context,  
a high capture rate is desired. We investigate the possibility of enhancing 
the electron capture rate by irradiating laser beam to ''atom''. The
 possibility of such enhancement can be understood as a consequence of an enhancement of the 
 electron wave function at the origin, $\Psi (0)$, through an increased effective
 mass of the electron. We find that an order of magnitude enhancement can be realized  
 by using a laser with intensity on the order of $10^{10}$ W/mm$^2$ and
 a photon energy on the order of $10^{-3}$ eV.
}
\maketitle
%
\section{Introduction}\label{intro}
Capture processes have a potential power to search for new physics.
For example, instead of neutrinoless double beta decay,
the process
\begin{eqnarray}
(Z,A) +e^-\rightarrow (Z-2,A)+e^+ ,
\end{eqnarray} 
where $(Z,A)$ denotes a nucleus of atomic number
$Z$ and mass number $A$, can be used to investigate Majorana
property of neutrinos.\cite{Bernabeu:1983yb,Ikeda:2005yh,Yoshimura:2005zq} The annihilation process of
muonium into two $\gamma$, i.e. $\mu+e\rightarrow 2\gamma$, 
is also promising in the search for lepton flavor violation.
\cite{Yshimura:Nufact05}
Furthermore, the neutrinos emerging from beta capture processes should provide a very
powerful source for oscillation experiments, because they have
a definite energy\cite{Sato:2005ma,Bernabeu:2005jh}.

The capture rate, $\Gamma$, is in general given by the form
\begin{eqnarray}
\Gamma = |\Psi (0)|^2 \int d\text{LIPS}|\mathcal{M}|^2,\label{cap_rate}
\end{eqnarray}
where $\Psi ({\bold{r}}) $ is the wave function of an electron \big[with $\Psi (0)$ that at the
center, i.e. a nucleus or muon\big] and 
$\int d\text{LIPS}=(2\pi)^4\delta^4(\sum_ip_i-\sum_f p_f)\prod_f dp^3_f/(2\pi)^32E_f$ is the Lorentz
invariant phase space (where the indices $i$ and $f$ represent the initial and final states), and  $\mathcal{M}$ is
the amplitude of the capture process with a plane 
wave electron. To earn a statistics, a rapid capture process
is desirable. The more rapid it is, the better an experiment will be.
Then how can we get more rapid process? We cannot control
the amplitude $\mathcal{M}$, since it is determined by the fundamental
physics. It is completely determined, at least in principle,
by the Lagrangian of the world. Thus, the only possibility for 
accelerating the process is to control the wave function of the 
electron.

In this paper, we show our attempt to control the wave function in such a manner 
that $|\Psi(0)|$ is larger than that in the uncontrolled case. That is, 
we show the possibility for obtaining a higher capture rate by
enhancing the wave function, since $\Gamma \propto |\Psi(0)|^2$.
The wave function is determined by the electron mass ,
$|\Psi(0)|^2 \propto m^3_{eff}$, where $m_{eff}$ is the effective mass
of the electron. Therefore, if we make the electron mass effectively
larger, we would realize a higher capture rate. In general, the wavelengths of a particle in a
medium and in vacuum will differ, hence, in terms of a particle, the mass of an electron
will be different in a medium than that in vacuum. This leads us to the idea that, in order to
increase the capture rate, we should immerse the system of atoms in a medium. 
For this purpose, the most effective medium is a photon bath.

With this intuition, we are led to consider the irradiation of a system with a laser. As above
mentioned, an electron is in a medium of photons when under irradiation. In terms of waves, the de
Broglie wavelength of the electron becomes {\it shorter} than that in the Coulomb electric field produced by the
nucleus owing to this irradiation. Hence, the Bohr radii of electrons become smaller due to the Bohr quantization
condition. In terms of particles, an electron propagates while being
scattered by the photons of the laser. Therefore, the electron must come nearer to the nucleus in
order to go
around it without falling into it. The common feature in both pictures is that the electron
goes around the nucleus in a closer orbit in the case that it is under irradiation.  
This implies that the binding energy of the electron
becomes larger. This fact can be interpreted as the electron gaining larger mass, because the binding
energy of an electron is proportional to its mass.
Thus, irradiation will result in a larger value of the wave function at the nucleus, and hence we can obtain a
higher capture rate. 
The idea of such irradiation with a laser
was proposed by Yoshimura\cite{Yoshimura:2005zq}. Indeed our idea
was largely inspired by his proposal. However, our result and the interpretation of the 
mechanism responsible for the enhancement are completely different from his idea.

In Sec.\ref{non_rel_lim}, we present a non-relativistic equation describing bound electrons 
in the presence of an external field, $\ev{eA(\vec{x},t)}\propto \sqrt{I/\omega^2}$, where
$\vec{A}(\vec{x},t)$ is the vector potential, and $I$ and $\omega$ are the intensity and the
angular frequency of the laser, respectively. Since $I$ and $\omega$ can be made lower as much as one
needs, we show that the competing ionization processes, namely, tunnel ionization and Multi-Photon
Ionization (MPI), can be suppressed. In Sec.\ref{sec_perturv}, we solve
the equation perturbatively. Then, we determine the dependences of the ratio of the capture
rate on the intensity and the photon energy in Sec.\ref{enhance_cap_rate}. 
We summarize and discuss our results in Sec.\ref{discussion}.  

\section{Non-relativistic Limit of the Dirac Equation} \label{non_rel_lim}
First, we derive a non-relativistic quantum equation with effective mass, $m_{eff}$, from the Dirac equation,
with a vector potential $\vec{A}(\vec{x},t)$ due to an irradiating laser and a Coulomb potential
$\phi(\vec{x})$. Throughout this paper, we use MKS units, and we employ $c$, $\hbar$
and $\epsilon_0$ to represent the speed
of light, Planck's constant and the permittivity of the vacuum, 
respectively. 
We express these constants explicitly. 

\subsection{The Dirac equation}
The Dirac equation with a vector potential and Coulomb potential is given by
\begin{align}
 i\hbar\del{}{t}\Psi(\vec{x},t)=\left[c\vec{\alpha}\cdot(\hat{\vec{p}}+e\vec{A}(\vec{x},t))
 +\beta m_ec^2-e\phi(\vec{x})\right]\Psi(\vec{x},t), \label{dirac-eq}
\end{align}
where $-e$ and $m_e$ are the electric charge and the mass of a bound
electron in vacuum, and $\vec{\alpha}$
and $\beta$ are given by
\begin{align}
 \vec{\alpha}=
 \begin{pmatrix}
  0 & \vec{\sigma} \\
  \vec{\sigma} & 0 
 \end{pmatrix}, \quad
 \beta=
 \begin{pmatrix}
  \vec{1} & 0 \\
  0 & -\vec{1}
 \end{pmatrix}
\end{align}
where $\vec{\sigma}$ represents the Pauli matrices. The Coulomb potential, $\phi$, is given by
\begin{align}
 \phi(\vec{x})=\frac{Ze}{4\pi\epsilon_0|\vec{x}|}.
\end{align}
The vector potential $\vec{A}(\vec{x},t)$ is given by the expectation
value of the photon field operator, $\hat{\vec{A}}(\vec{x},t)$, with respect to a coherent 
state of photons representing the laser light.\cite{Loudon} In this paper, we consider the situation
in which two identical lasers 
irradiate the atom from opposite sides. Such a situation can be realized experimentally by placing the atom in a box
whose sides are mirrors and irradiating it in this environment. The vector potential in this case
takes the form 
\begin{align}
 \vec{A}(\vec{x},t)
 =2\sqrt{\frac{2\hbar N}{\epsilon_0 \omega}}\cos(\vec{k}\cdot\vec{x})
 \cos(\omega t+\phi_\alpha) \vec{a}_{\vec{k}}, \label{vec-pot} 
\end{align}
where $N$ is the average photon number density and $\vec{a}_{\vec{k}}$ is the polarization vector. 
The quantities, $\vec{k}$, $\hbar\omega$ $(=E_\gamma)$ and $\phi_{\alpha}$ are the wavevector, energy per photon
and initial phase, respectively. Note that in Eq.\eqref{dirac-eq}, the external vector potential,
$\vec{A}(\vec{x},t)$, appears in the form $\hat{\vec{p}}+e\vec{A}(\vec{x},t)$. Hence it modifies the
momentum or wavelength. As shown in Sec.\ref{sec_non-rel-lim}, this modification can be interpreted
as a modification of the mass.

\subsection{Effective Mass}
Following the intuitive understanding described in Sec.\ref{intro}, we incorporate the vector
potential into the scalar mass to obtain the effective mass. 

Let $U$ be the diagonalization matrix and $m_{eff}$ be the eigenvalue of 
$ce\vec{\alpha}\cdot\vec{A}+\beta m_e c^2$. Then, after a simple calculation, we obtain
\begin{align}
 m_{eff}(\vec{x},t)&=\sqrt{m^2_e+\big(\frac{e}{c}\vec{A}(\vec{x},t)\big)^2},\label{eff_mass} \\
 U(\vec{x},t)&=
\begin{pmatrix}
 \vec{1}\cos\Theta(\vec{x},t) & -\vec{\sigma}\cdot\vec{a}(\vec{x},t)\sin\Theta(\vec{x},t) \\
 \vec{\sigma}\cdot\vec{a}(\vec{x},t)\sin\Theta(\vec{x},t) & \vec{1}\cos\Theta(\vec{x},t)
\end{pmatrix},\\
\vec{a}(\vec{x},t) &\equiv \frac{\vec{A}(\vec{x},t)}{|\vec{A}(\vec{x},t)|},\\
\sin\Theta(\vec{x},t)&=\frac{(m_{eff}(\vec{x},t)-m_e)c^2}{\sqrt{\big(ce\vec{A}(\vec{x},t)\big)^2 +(m_{eff}(\vec{x},t)-m_e)^2c^4}},\\
\cos\Theta(\vec{x},t)&=\frac{|ce\vec{A}(\vec{x},t)|}{\sqrt{\big(ce\vec{A}(\vec{x},t)\big)^2+(m_{eff}(\vec{x},t)-m_e)^2c^4}}.
\end{align}
Here $\Theta$ is the mixing angle between the vacuum mass eigenstates
and the  effective mass eigenstates. Here we define $m_{eff}$ as the effective mass of
electrons in the photon medium. Note that the direct effect of irradiation by the laser is to modify
the momentum 
of a bound electron, not its mass. This modification results in a change of the binding
energy. Our idea is that this modification of the momentum can be represented as a modification of the
mass by putting the vector potential into the mass, which also results in a change of the energy. 
From eq.\eqref{eff_mass}, we see that $m_{eff}$ becomes larger than $m_e$ in the presence of the
external field. Because the atomic electron capture rate is proportional to $|\Psi(0)|^2\propto
m_{eff}^3$, high intensity laser irradiation can enhance the capture rate.  

Following the above discussion, the Dirac equation \eqref{dirac-eq} is rewritten as follows \big[For simplicity, we denote
$\vec{A}(\vec{x},t)$, $m_{eff}(\vec{x},t)$ and $U(\vec{x},t)$ as $\vec{A}$, $m_{eff}$ and $U$.\big]
\begin{align}
 i\hbar \del{}{t}\Phi 
 =\left[ cU^\dagger \vec{\alpha} U\cdot \hat{\vec{p}}+\beta m_{eff}c^2-e\phi \right]\Phi 
 -U^\dagger\left\{i\hbar \del{}{t}U
 -c\alpha\cdot(\hat{\vec{p}}U)\right\}\Phi. \label{red_dirac_eq}
\end{align}
Here we define new mass eigenstates as $\Phi\equiv U^\dagger\Psi$. 
Note that if $|e\vec{A}|$ is much smaller than 
$m_ec$, then $m_{eff}$ becomes $m_e\left(1+e^2\vec{A}^2/2m_e^2 c^2\right)$. This can be interpreted as
a shift of the energy reference. The effective mass can be also expressed in terms of  the Keldysh parameter,
$\kappa$ :
\begin{align}
 m_{eff}&=m_e\sqrt{1+\left(\frac{2\alpha}{\kappa}\right)^2}, \\
 \kappa&=\sqrt{\frac{m_e^2 \alpha E_\gamma}{8\pi (\hbar c)^2\hbar I}}.
\end{align}
Here $\alpha=e^2/4\pi \epsilon_0\hbar c$ is the fine structure constant.

In the following, we investigate the case in which $E_\gamma$ is 
 $\mathcal{O}(10^{-3})$ eV or less and hence the wavelength of the laser light is
 $\mathcal{O}(10^{-6})$ m or longer, which is 
$10^5$ times greater than the atomic size. The period of the laser light, $T_{\text{Laser}}$, in this region is
longer than the average time, $T_{e}$, for an electron to orbit the nucleus; 
explicitly, we have $T_{\text{Laser}}= 4.1\times 10^{-15}/E_\gamma$[eV] (s) 
$>2.9\times 10^{-16}=T_e$ (s). These facts suggest that the dependence of $U$
on the position and the time is very weak. Therefore, we can ignore the
last term in Eq. (\ref{red_dirac_eq}) and treat $m_{eff}$ as a constant at each time. 
Then the Dirac equation \eqref{red_dirac_eq} takes the simple form 
\begin{align}
 i\hbar\del{}{t}\Phi(\vec{x},t)=\hat{H}\Phi(\vec{x},t), \label{eff-dirac-eq}
\end{align}
where
\begin{align}
 \hat{H}&= c\vec{\alpha}\cdot\hat{\vec{p}}
 +c\left\{-(\vec{\alpha}\cdot\vec{a})+\beta\sin2\Theta 
 + (\vec{\alpha}\cdot\vec{a})\cos2\Theta\right\}(\vec{a}\cdot\hat{\vec{p}})\nn \\
 &\quad +\beta m_{eff}c^2 -e\phi(\vec{x}).
\end{align}
Here we have used the relation
\begin{align}
 U^\dagger \vec{\alpha}U=\vec{\alpha}-(\vec{\alpha}\cdot\vec{a})\vec{a}
  +\bigg(\beta\sin2\Theta+(\vec{\alpha}\cdot\vec{a})\cos2\Theta\bigg)\vec{a}.
\end{align}

We make the further approximation that under the static assumption that $T_{\text{Laser}}$ is
sufficiently larger than $T_e$, the vector potential,
Eq. \eqref{vec-pot}, can be approximated as
\begin{align}
  \vec{A}^2(\vec{x},t)=\frac{4\hbar N}{\epsilon_0 \omega}.\label{static_A}
\end{align}
As mentioned above, the
effective mass, Eq. \eqref{eff_mass}, is determined by the vector potential at each time. For our purpose,
$m_{eff}^3$ should be replaced by its time average, because the capture rate is proportional to
$m_{eff}^3$. But considering the time period of interest here, there is only a few percent difference
between the time average of $m_{eff}^3$ and that obtained with $\vec{A}^2(\vec{x},t)$ replaced by
Eq.\eqref{static_A}. Furthermore, this replacement is often made in the field of laser science, and
it makes the form of the effective mass easy to understand. 
For these reasons, we simply replace $\vec{A}^2(\vec{x},t)$ in the effective mass by the
quantity in Eq.\eqref{static_A} in the following discussion. 

The average photon number density $N$ is related to the laser intensity $I$ as
\begin{align}
I\simeq c\hbar \omega N. \label{int_num}
\end{align}
Thus, the vector potential can be expressed in terms of $I$ as
\begin{align}
 \vec{A}^2(\vec{x},t)\simeq \frac{4I}{\epsilon_0~\omega^2c}. \label{vec-pot-int}
\end{align}
We note that the vector potential is proportional to $\sqrt{I}/E_\gamma$.
 There are two competing ionization processes that must be suppressed in order for the
enhancement mechanism to sufficiently work, tunnel ionization and multi-photon ionization
(MPI). The probability for tunnel ionization\cite{Ammosov} is proportional to 
$\exp(-\mathcal{E}_{\text{C}}/\mathcal{E}_{\text{L}})$, where $\mathcal{E}_{\text{C}}$ is the
strength of the Coulomb electric field produced by the nucleus, and $\mathcal{E}_{\text{L}}$ is that 
produced by the laser. Because $\mathcal{E}_{\text{L}}$ is given by $\sqrt{I/\epsilon_0 c}$, 
$\mathcal{E}_{\text{L}}$ becomes smaller than $\mathcal{E}_{\text{C}}$ in the hydrogen atom for
$I<10^{13}$ W/mm$^2$. Therefore, tunnel ionization can be ignored for  
$I\ll 10^{13}$ W/mm$^2$. For instance, if $I$ is $10^{10}$ W/mm$^2$ and $\mathcal{E}_{\text{C}}$ is 
$5.0 \times 10^{11}$ V/m, we have 
$\exp(-\mathcal{E}_{\text{C}}/\mathcal{E}_{\text{L}})\sim 10^{-8\times 10^{4}}$, and hence tunnel
 ionization is very greatly suppressed in this case. 
Furthermore, it is believed that there are no resonances of photon 
 absorption in MPI cross sections when $E_\gamma$ is smaller sufficiently than the difference in the
 binding energies of the $1s$ and $2s$ states,  
$\Delta E_{\text{bin}}=E_{2s}-E_{1s}$. Note that electrons in $1s$ and, at most, $2s$ play roles in
 capture processes. In this case, the cross sections are proportional to 
$(N\alpha)^{E_{\text{bin}}/E_\gamma} \sim(I\alpha/E_\gamma)^{E_{\text{bin}}/E_\gamma}$ and 
go to zero as $E_\gamma \rightarrow 0$ for fixed $I/E_\gamma$. The same conclusion is derived from 
consideration of the Keldysh parameter, and these facts indicate that MPI is even more suppressed
 than tunnel ionization when $\kappa<1$.
For the case with $I=10^{10}$ W/mm$^2$ and $E_\gamma=10^{-3}$ eV, $\kappa$ is
 $3.8\times 10^{-3}$, and MPI is suppressed.  
We then conclude that the ionizing processes can be suppressed by using a low energy, low intensity laser. 
Contrastingly, $m_{eff}$ is an increasing function of $I/E_\gamma^2$, or $1/\kappa$. Therefore, we can 
simultaneously increase the effective mass and suppress the ionization processes by arbitrarily
 large amount, even if $I$ and $I/E_\gamma$ are fixed at small values. For this reason, we can
 ignore tunnel ionization and MPI for $1s$
 and $2s$ states. 

\subsection{Non-relativistic Limit}\label{sec_non-rel-lim}
Since binding energies of atomic electrons are very small compared to their rest mass, a non-relativistic
form of the Dirac equation gives a good approximation. Therefore, we reduce Eq.\eqref{eff-dirac-eq} to its
non-relativistic form.

To take the non relativistic limit, we separate the rest mass from the wave function :
\begin{align}
 \Phi(\vec{x},t)&=\exp\left(-\frac{im_{eff}c^2}{\hbar}t\right)\varphi_{NR}(\vec{x},t), \nn\\
 \varphi_{NR}(\vec{x},t)&=
\begin{pmatrix}
 \varphi(\vec{x},t) \\
 \eta(\vec{x},t)
\end{pmatrix}.
\end{align}
Next, inserting the above relations into Eq.\eqref{eff-dirac-eq}, we have
\begin{subequations}
\begin{align}
 i\hbar\del{}{t}\varphi &= \bigg(-e\phi+c\tsin (\vec{a}\cdot\hat{\vec{p}})\bigg)\varphi
  +\bigg(c\vec{\sigma}\cdot\hat{\vec{p}}
  +c(\tcos-1)(\vec{a}\cdot\hat{\vec{p}})(\vec{\sigma}\cdot\vec{a})\bigg)\eta, \\
 i\hbar\del{}{t}\eta &= \bigg(-e\phi-c\tsin (\vec{a}\cdot\hat{\vec{p}})-2m_{eff}c^2\bigg)\eta
  +\bigg(c\vec{\sigma}\cdot\hat{\vec{p}}
  +c(\tcos-1)(\vec{a}\cdot\hat{\vec{p}})(\vec{\sigma}\cdot\vec{a})\bigg)\varphi,
\end{align}
\end{subequations}
where $\tsin=\sin2\Theta$ and $\tcos=\cos2\Theta$.
Since $|\vec{p}|/m_{eff}c \ll 1$, the second equation reduces to
\begin{align}
 \eta=\frac{1}{2m_{eff}c}\left[\vec{\sigma}\cdot\hat{\vec{p}} + (\tcos-1) (\vec{\sigma}\cdot\vec{a})
 (\vec{a}\cdot\hat{\vec{p}})\right]\varphi.
\end{align}
Then, inserting the  above result into the first equation, we obtain the non-relativistic equation :
\begin{align}
 i\hbar\del{}{t}\varphi(\vec{x},t)&=
 \left[\frac{1}{2m_{eff}}\left(\hat{\vec{p}}^2-\tsin^2(\vec{a}\cdot\hat{\vec{p}})^2\right)
  +c\tsin(\vec{a}\cdot\hat{\vec{p}})-e\phi(x)\right]\varphi(\vec{x},t).  \label{eff-dirac-eq2}
\end{align}
Now it is seen that the mass of the electron has been replaced by $m_{eff}$, and the
vector potential has disappeared in the kinetic term. Thus, the modification of the momentum can be
interpreted as a modification of the mass.
\section{Perturbative Expansion}\label{sec_perturv}
In the previous section, we derived the non-relativistic equation with
the effective mass, Eq.\eqref{eff-dirac-eq2}. We now show that the probability for finding the electron at
the origin is mainly determined by the zeroth-order wave function, which is proportional to
$m_{eff}^{3}$. Therefore, a high intensity laser can enhance the capture rate of the atomic decay. 
To show this, we solve the equation perturbatively up to second order. 

First, we rewrite the Hamiltonian appearing in
Eq.\eqref{eff-dirac-eq2} in a simpler 
form. Let us take $\vec{A}=(0,0,A)$ and complete the square involving $\hat{p}_z$ 
\begin{align}
 \frac{1}{2m_{eff}}\left(\hat{p}_z^2 -\tsin^2\hat{p}_z^2 \right)+c\tsin\hat{p}_z
 =\frac{\tcos^2}{2m_{eff}}\left(\hat{p}_z+m_{eff}c\frac{\tsin}{\tcos^2}\right)^2-\frac{1}{2}m_{eff}c^2\tilde{t}^2.
 \label{non-rel-kinetic}
\end{align}
Here, we have $\tilde{t}=\tan2\Theta=\sin2\Theta/\cos2\Theta$. The first term in Eq.\eqref{non-rel-kinetic}
is reduced to $\hat{p}_z^2$ if $\varphi(\vec{x},t)$ is replaced with 
$\exp(-im_{eff}c\tsin z/\hbar\tcos^2)\varphi(\vec{x},t)$, and the second term is dropped, because it is just
like vacuum energy. Furthermore, we separate the time dependence. Thus we obtain the equation that
we hope to solve,
\begin{subequations}
\begin{align}
 E\varphi(\vec{x})&=(\hat{H}_0+\hat{H}_{int})\varphi(\vec{x}),\\
 \hat{H}_0&=\frac{\hat{\vec{p}}^2}{2m_{eff}}-e\phi(x),\\
 \hat{H}_{int}&=-\frac{\tsin^2}{2m_{eff}}\hat{p}_z^2. \label{h_int}
\end{align}
\end{subequations}

In the following, we calculate the energy and wave functions of a hydrogen-like $1s$ bound state of
an atom with atomic number $Z$.

\subsection{Zeroth Order}
The zeroth-order energies and wave functions are eigenvalues and the eigenfunctions of
$\hat{H}_0$. These are given by
\begin{align}
 \varphi^{(0)}_{nlm}(\vec{x})&=R_{nl}(r)Y_{lm}(\theta,\phi),
\end{align}
where $n,~l$ and $m$ are the principal quantum number, azimuthal quantum number and magnetic quantum
number. The functions $R_{nl}(r)$ and $Y_{lm}(\theta,\phi)$ are listed in Appendix \ref{0th-wavefunc}.

The energy of zeroth order is 
\begin{align}
  E_n^{(0)}&=-\frac{\hbar^2}{2m_{eff}a_1^2}\frac{1}{n^2}, 
\end{align}
where $a_1$ is the orbital radius of the electron,
\begin{align}
 a_1&=\frac{4\pi\epsilon_0\hbar^2}{Zm_{eff}e^2}.
\end{align}

\subsection{First Order}

The corrections to the energy and wave function of the $1$s state at first order are 
\begin{align}
 E^{(1)}_1=\frac{\tsin^2}{3}E^{(0)}_1 \label{energy_1st}
\end{align}
and 
\begin{align}
 \varphi^{(1)}_{100}(\vec{x})
  =\tsin^2\sum_{n=2}f^{(1)}_{n00}\varphi^{(0)}_{n00}
  +\tsin^2\sum_{n=3}f^{(1)}_{n20}\varphi^{(0)}_{n20} \label{first-order-wavefunc},
\end{align}
where 
\begin{subequations}
\begin{align}
 f^{(1)}_{n00}&=\frac{8}{3}\frac{n^{5/2}}{(n^2-1)^2}\left(\frac{n-1}{n+1}\right)^{n-1},\\
 f^{(1)}_{n20}&=\frac{\sqrt{5}}{15}\frac{n^2}{n^2-1}\left(\frac{2}{n+1}\right)^5
  \sqrt{\frac{(n-3)!}{(n+2)!}}\nn\\
  &\quad \times\sum_{r=0}^{n-3}\comb{n+2}{n-3-r}\frac{(r+3)!}{r!}\big(n(r+5)+1\big).
\end{align}\label{weight_1st}
\end{subequations}

\subsection{Second Order}
%

The corrections to the energy and wave function of the $1$s state at second order are 
\begin{align}
 E^{(2)}_1=\tsin^4 E^{(0)}_1
  \left[
   \frac{3}{4} {f^{(1)}_{200}}^2
   +\sum_{n=3}\frac{n^2-1}{n^2}\left({f^{(1)}_{n00}}^2+{f^{(1)}_{n20}}^2\right)
  \right] \label{energy_2nd}
\end{align}
and 
\begin{align}
 \varphi^{(2)}_{100}(\vec{x})=\sum_{n\neq 1 \atop lm}
  \tsin^4 f^{(2)}_{nlm}\varphi^{(0)}_{nlm}
  +\tsin^4C^{(2)}\varphi^{(0)}_{100},
\end{align}
where $C^{(2)}$ is 
\begin{align}
  C^{(2)}&=-\frac{1}{2}\sum_{n=2}{f^{(1)}_{n00}}^2
  -\frac{1}{2}\sum_{n=3}{f^{(1)}_{n20}}^2
\end{align}
and $f^{(2)}_{nlm}$ is given by the following :
\begin{subequations}
\begin{align}
 f^{(2)}_{n00}&=-\frac{1}{3}\frac{n^2}{n^2-1}f^{(1)}_{n00}
  +\frac{n^2}{n^2-1}\sum_{n'=2}f^{(1)}_{n'00}N^{00}_{nn'}
   \left(\frac{1}{3n'}F^{00}_{nn'}(2)+\frac{2}{3}F^{00}_{nn'}(1)\right)\nn\\
  &+\frac{2\sqrt{5}}{15}\frac{n^2}{n^2-1}\sum_{n'=3}f^{(1)}_{n'20}N^{02}_{nn'}
   \left( -\frac{1}{{n'}^2}F^{02}_{nn'}(2)+\left(\frac{3}{n'}+2\right)F^{02}_{nn'}(1)\right. ,\\
 &\hspace{4cm}+7(n'-3)F^{02}_{nn'}(0)-7(n'+2)G^{02}_{nn'}(0) \bigg),\nn\\
 f^{(2)}_{n20}&=\left(C^{(1)}-\frac{1}{3}\frac{n^2}{n^2-1}\right)f^{(1)}_{n20}\nn\\
  &+\frac{2\sqrt{5}}{15}\frac{n^2}{n^2-1} \sum_{n'=2}f^{(1)}_{n'00} N^{20}_{nn'} 
   \left( -\frac{1}{{n'}^2}F^{20}_{nn'}(2)-\left( \frac{3}{n'}-2 \right)F^{20}_{nn'}(1)\right.\nn\\
  &\hspace{4cm} +3(n'-1)F^{20}_{nn'}(0)-3n'G^{20}_{nn'}(2) \bigg) \nn \\
  &+\frac{1}{7}\frac{n^2}{n^2-1}\sum_{n'=3}f^{(1)}_{n'20}N^{22}_{nn'}
   \left(-\frac{11}{3n'}F^{22}_{nn'}(2)+\frac{22}{3}F^{22}_{nn'}(1)\right),\\
 f^{(2)}_{n40}&=\frac{4\sqrt{5}}{35}\frac{n^2}{n^2-1}\sum_{n'=3}f^{(1)}_{n'20}N^{42}_{nn'}
  \left(-\frac{1}{{n'}^2}F^{42}_{nn'}(2)-\left(\frac{7}{n'}-2\right)F^{42}_{nn'}(1)
  \right.\nn\\
 &\hspace{4cm} +7(n'-3)F^{42}_{nn'}(0)-7(n'+2)G^{42}_{nn'}(0)\bigg),
\end{align}\label{weight_2nd}
\end{subequations}
with $f^{(2)}_{nlm}=0$ for all other values of $l$ and $m$. The coefficients $N^{ll'}_{nn'}$, $F^{ll'}_{nn'}(p)$ and
$G^{ll'}_{nn'}(p)$ are given in Appendix \ref{formulae}.


Note that in the perturbative corrections, only states with even values of $l$ and $m=0$ appear. The
situation is the same at higher orders. This can be understood from the fact that parity is
conserved in Eq.\eqref{eff-dirac-eq2}, and the magnetic quantum number is also conserved since there
remains rotational symmetry about the z axis. 

\begin{figure}[ht]
\begin{center}
\begin{tabular}{cc}
\includegraphics[height=55mm]{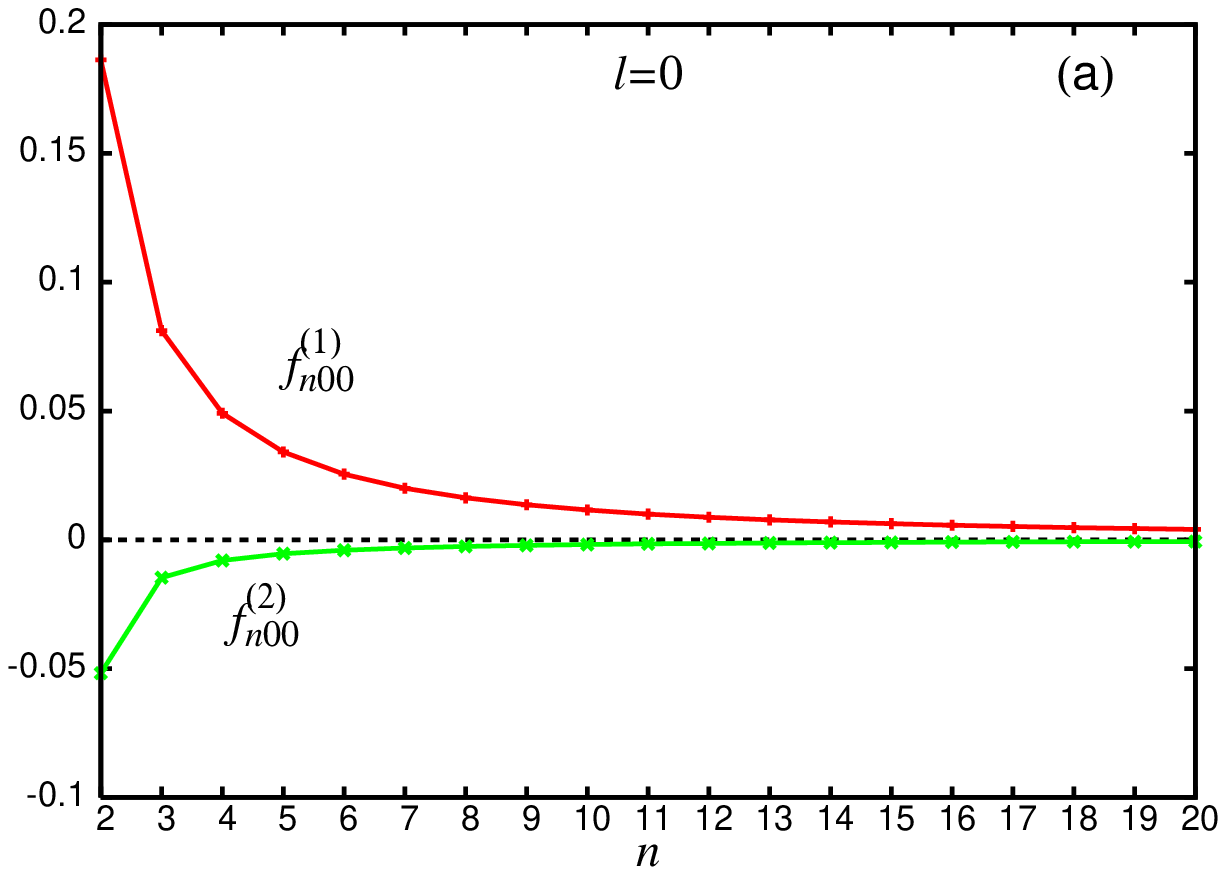} & 
\includegraphics[height=55mm]{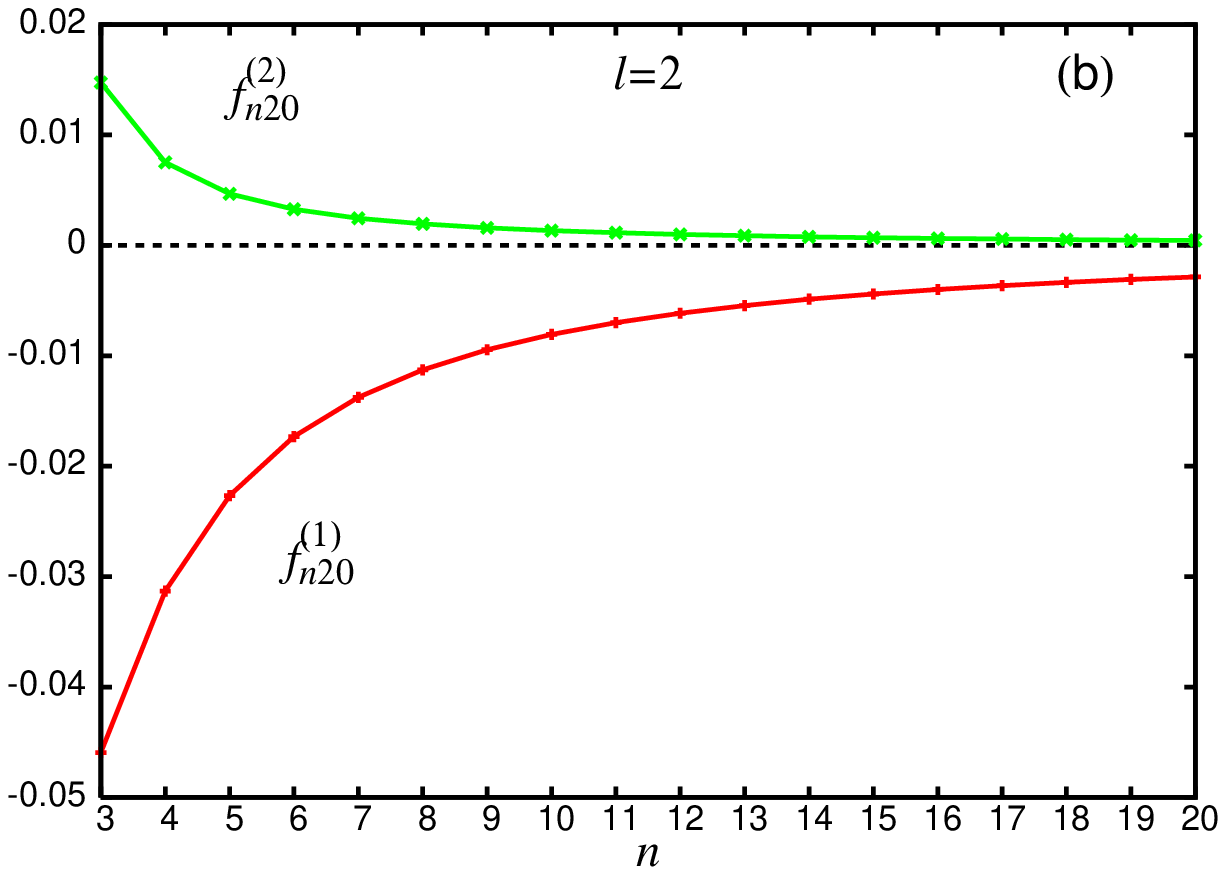} \\
\includegraphics[height=55mm]{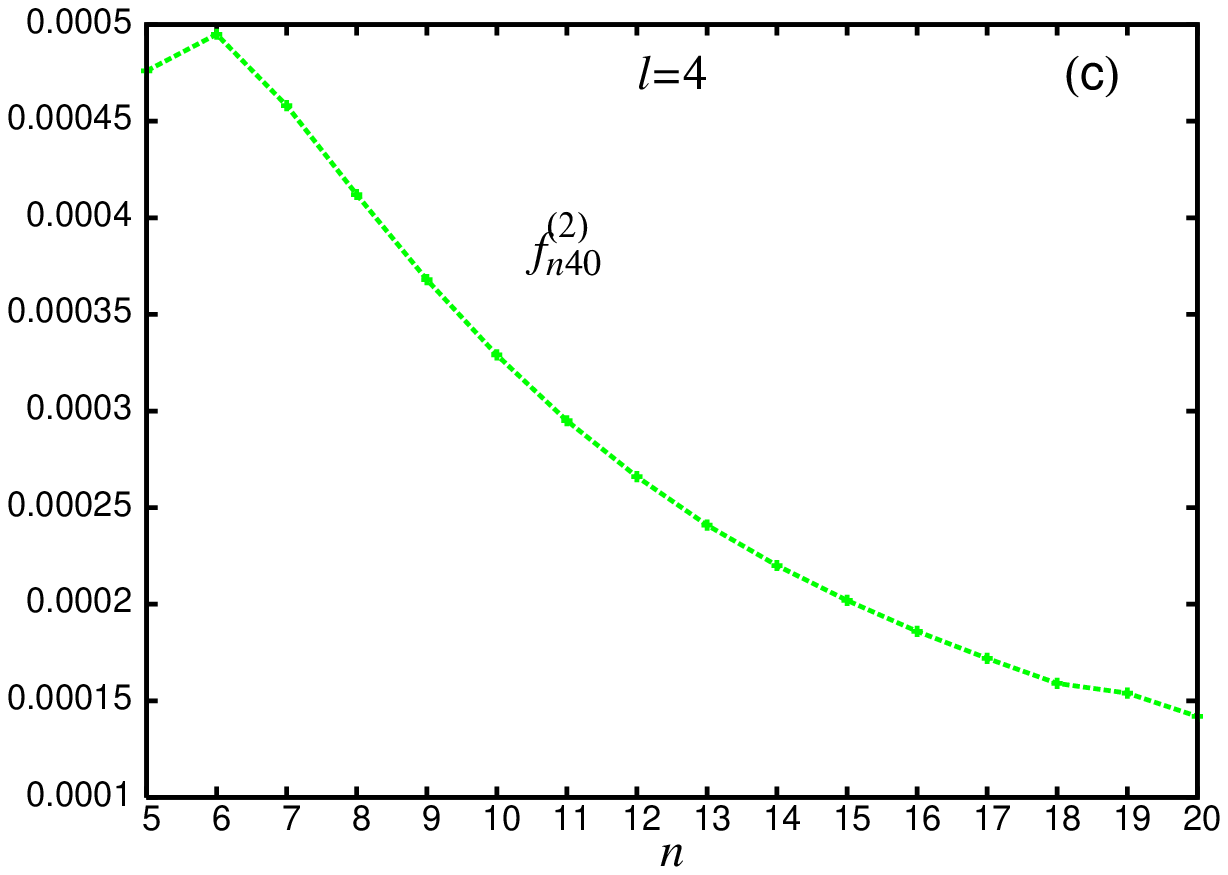} & 
\includegraphics[height=55mm]{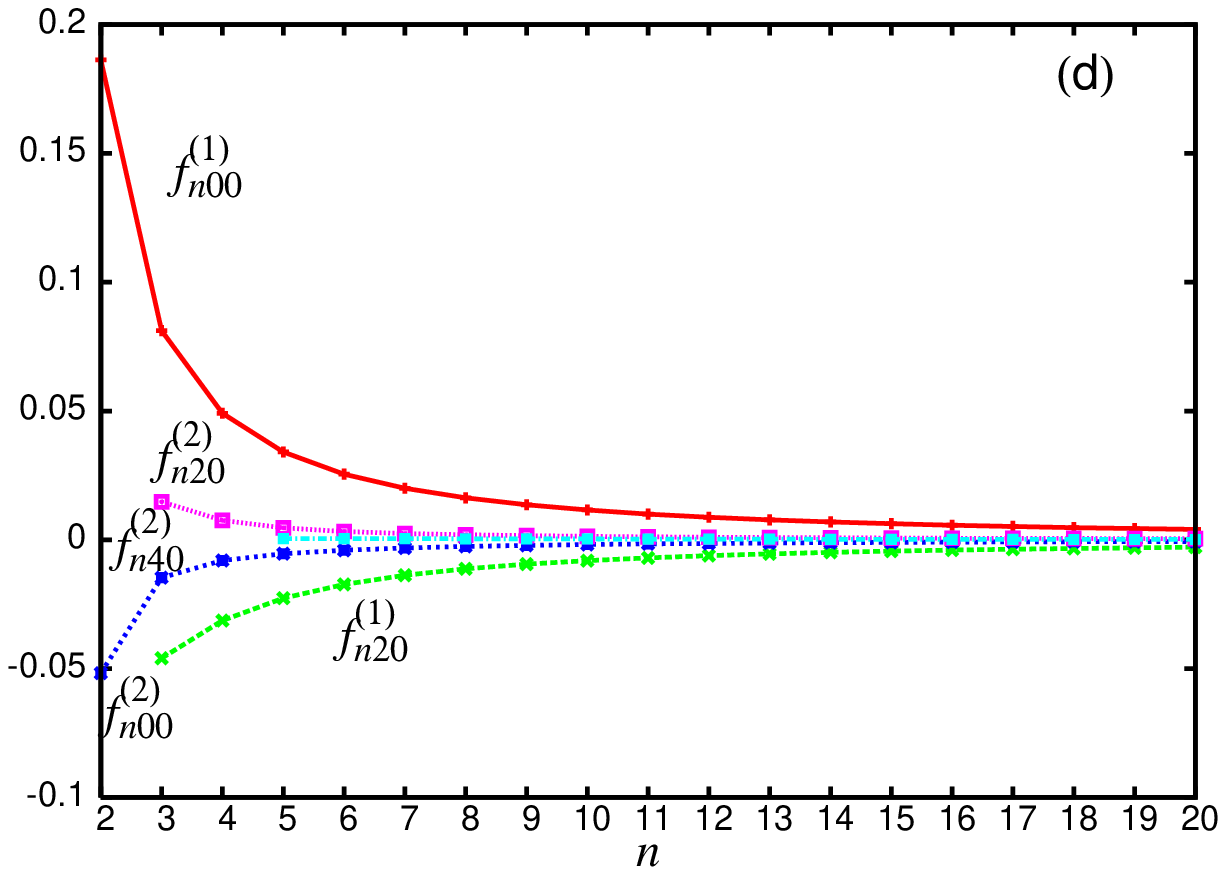} 
\end{tabular}
\caption{The graphs in (a), (b) and (c) plot the coefficients for $l=0$, $l=2$ and $l=4$ to $n$,
 respectively. The graph in (d) plots all coefficients in Eqs. \eqref{weight_1st} and \eqref{weight_2nd}.}
\label{weights}
\end{center}
\end{figure}

In Fig \ref{weights}, we plot the coefficients of the wave functions, given in Eqs.\eqref{weight_1st} and
\eqref{weight_2nd}, for each principal quantum number. 
As seen from the figures, the first-order and second-order corrections are much smaller than $1$ (the zeroth
order value) and go to zero in the limit $n\rightarrow \infty$. It is also
seen that the first-order
corrections are larger than the second-order corrections. We thus find that the probability at the origin
is mainly determined by the zeroth-order wave function.

\section{Enhancement of the Atomic Electron Capture Rate} \label{enhance_cap_rate}
In this section, we present the results of numerical calculations of the atomic electron capture
rate, Eq.\eqref{cap_rate}, mixing angle, 
Eq.\eqref{energy_1st} and
energies, Eq.\eqref{energy_2nd}.

From Eq.\eqref{cap_rate} and the relation $|\Psi(0)|^2 \simeq |\varphi(0)|^2$, the ratio of the
capture rates with and without the irradiation is given by
\begin{align}
 \frac{\Gamma_{\text{laser}}}{\Gamma}=\frac{|\Psi_{\text{laser}}(0)|^2}{|\Psi(0)|^2}
  \propto \left(\frac{m_{eff}}{m_e}\right)^3,\label{ratio_cap_rate}
\end{align}
where $m_{eff}$ is a function of the intensity given in Eq.\eqref{vec-pot-int}.

\begin{figure}[ht]
\begin{center}
\begin{tabular}{c@{\hspace{6mm}}c}
\includegraphics[height=60mm]{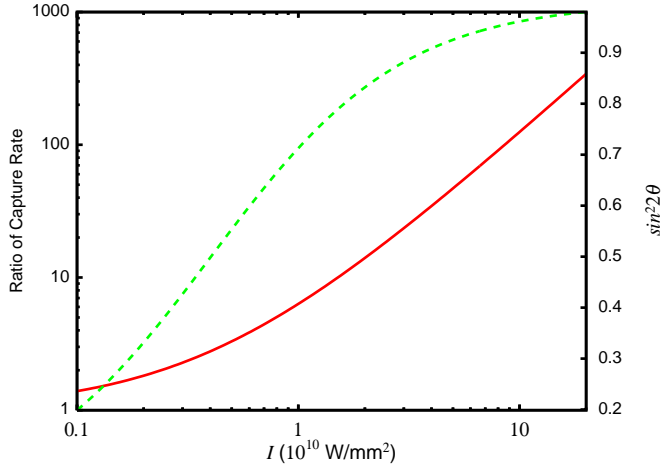} 
\end{tabular}
\caption{Ratio of the atomic electron capture rates and
$\sin^22\Theta$ under laser irradiation for $E_\gamma=10^{-3}$ eV plotted as a function of
 $I$. The laser intensity is varied from $10^{9}$ to $2\times 10^{11}$ W/mm$^2$. 
The red solid curve represents the ratio of the capture
 rates, and the green dashed curve is $\sin^22\Theta$. The left horizontal axis
 is for the ratio of the capture rates, and the right axis is for $\sin^22\Theta$.}
\label{effmass_mixing}
\end{center}
\end{figure}

As discussed in Sec.\ref{non_rel_lim}, we take $E_\gamma=10^{-3}$ eV and $I=10^{10}$ W/mm$^2$ as the reference
values so that the MPI and tunnel ionization can be ignored.
Figure \ref{effmass_mixing} shows the ratio of the capture rates, Eq.\eqref{ratio_cap_rate}, and 
$\sin^22\Theta$ at $E_\gamma=10^{-3}$ eV. For the data plotted in all figures, we have $Z=1$, $m_e=0.5$ MeV and $4\pi\epsilon_0\hbar c/e^2=137$. The intensity is varied from $I=10^{9}$ to $2\times 10^{11}$ W/mm$^2$.
It is seen that the ratio of the capture rates and the mixing angle grow
 as the laser intensity increases. It is also seen that at 
$I=10^{10}$ W/mm$^2$, $\sin^22\Theta$ is approximately $0.7$, and the ratio is approximately $6$
 times larger than that without irradiation. 

\begin{figure}[ht]
\begin{center}
\begin{tabular}{c@{\hspace{6mm}}c}
\includegraphics[height=55mm]{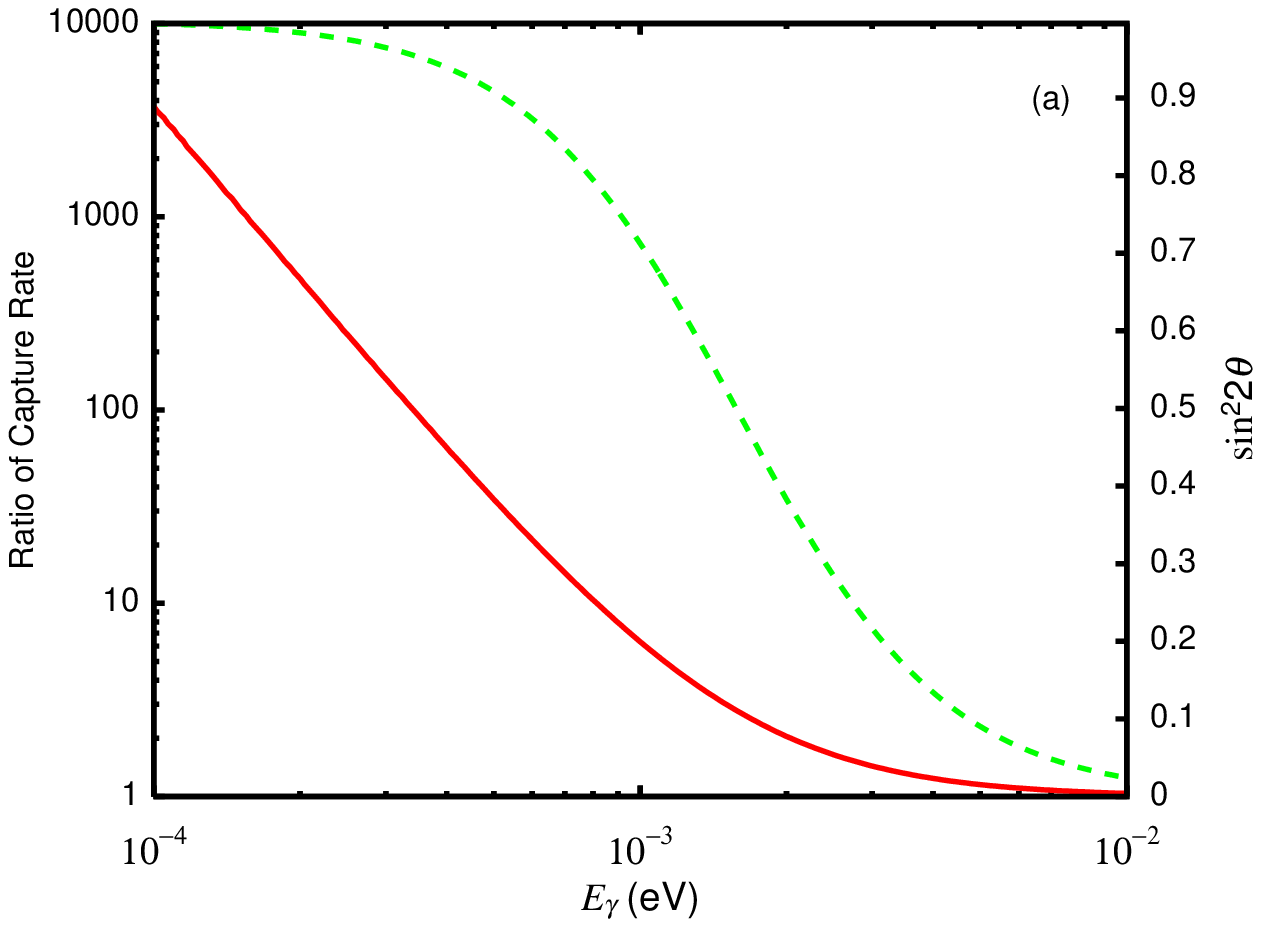} &
\includegraphics[height=55mm]{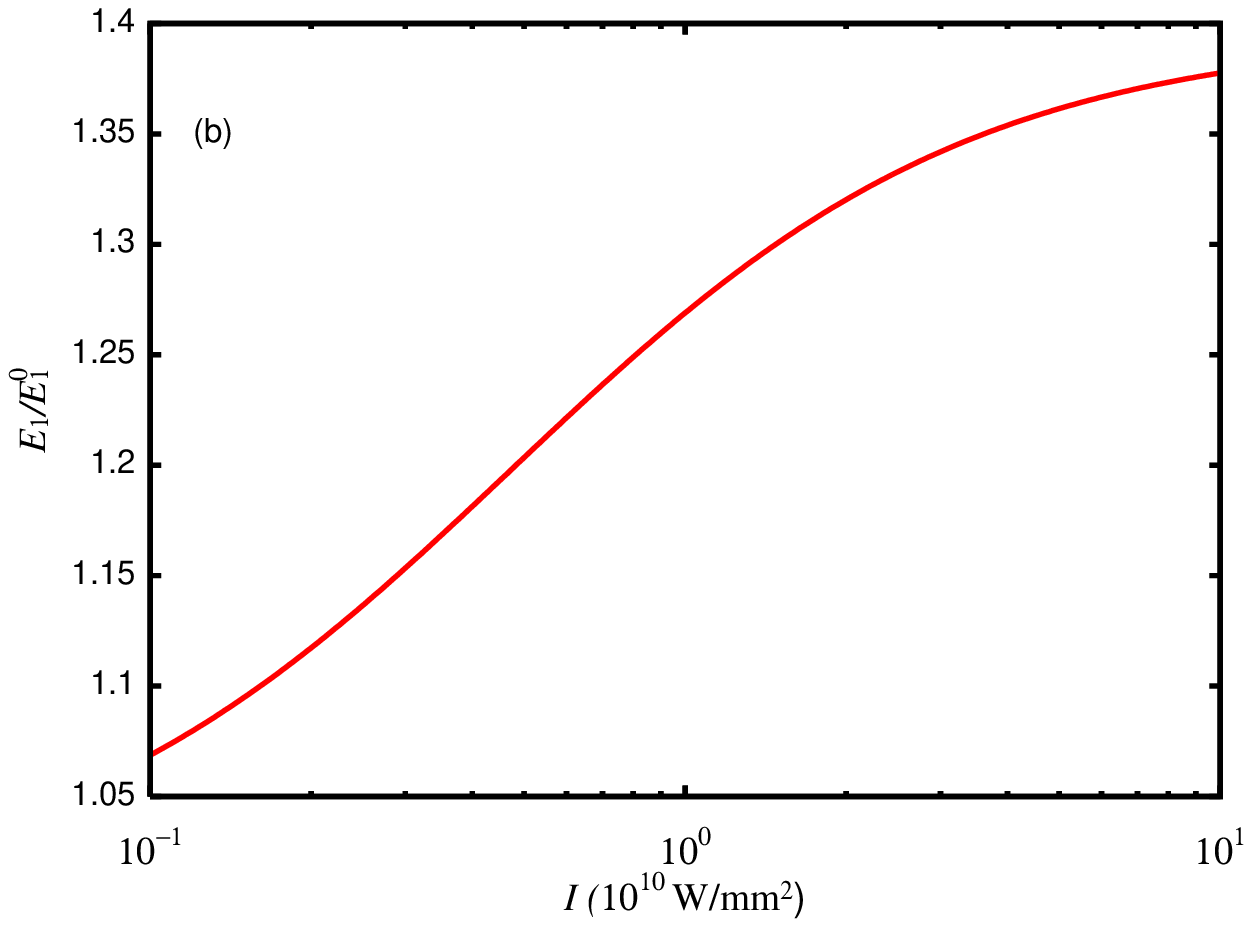} 
\end{tabular}
\caption{The graph in (a) shows the $E_\gamma$ dependence of the ratio of the capture rates at $I=10^{10}$
 W/mm$^2$. There, the red solid curve represents the ratio of the capture rates, and the green dashed curve the
 mixing angle. The graph in (b) shows the intensity dependence of the energy, Eq.\eqref{energy_ratio}, at $E_\gamma=10^{-3}$ eV.}
\label{energy_dep}
\end{center}
\end{figure}

Figure \ref{energy_dep}(a) shows the $E_\gamma$ dependence of the ratio
of capture rates at the intensity
$10^{10}$ W/mm$^2$. Figure \ref{energy_dep}(b) plots the ratio of the energy to the zeroth-order
energy as a function of the intensity at $E_\gamma=10^{-3}$ eV, 
\begin{align}
 \frac{E_1^{(0)}+E_1^{(1)}+E_1^{(2)}}{E_1^{(0)}}.\label{energy_ratio}
\end{align}

It is seen in Fig.\ref{energy_dep}(a) that the capture ratio and mixing angle become smaller as the
photon energy becomes larger. Thus we see that a lower energy photon is suitable for the enhancement. This can be
understood in terms of the average photon number within the orbital radius, as follows. From
Eq.\eqref{int_num}, the average photon number within a radius of  $4\pi a_1^3/3$ at $I=10^{10}$ W/mm$^2$ is
given by
\begin{align}
 2 N \times \frac{4\pi}{3}a_1^3&=\frac{4I}{cE_\gamma}\frac{4\pi}{3}a_1^3
  \sim \frac{4.2\times 10^{-4}}{E_\gamma\text{[eV]}}, \label{photon_num}
\end{align}
where the factor of $2$ reflects the fact that two identical lasers irradiate the atom from
opposite sizes. The photon bath becomes denser as the photon energy becomes lower. Thus,
the capture rate increases as the energy of the laser decreases.
From Fig.\ref{energy_dep}(b), we can see that the energy of a bound electron increases by approximately $27$\%
at $10^{10}$ W/mm$^2$. This increase comes mainly from the first-order correction, Eq.\eqref{energy_1st}. we interpret
this behavior as resulting from the electron being bound more closely to the nucleus or the muon for
higher intensity laser light.

\section{Summary and Discussion}\label{discussion}
We have studied the possibility to enhance the atomic electron capture
rate by employing laser irradiation. In this situation, electrons are immersed in a medium
consisting of photons. When photons interact with electrons efficiently,
electrons are difficult to move. The effect of the laser is expressed by a
vector potential. Therefore, it appears as a modification of the electron
momentum, which, in the non-relativistic case, can be interpreted as a
change of its mass. We reduced the Dirac equation including vector and
Coulomb potentials, (\ref{dirac-eq}), to a non-relativistic equation
with effective mass $m_{eff}$, Eq.(\ref{eff-dirac-eq2}).
The effective mass was, indeed, found to be heavier than that in vacuum.

We solved this equation up to second order in the perturbation,
$\hat{H}_{int}$, given in (\ref{h_int}), in Sec.\ref{sec_perturv}. Though 
the perturbative term appears to be large, we showed that it can be treated as a 
perturbation.  The validity of the perturbative calculations is supported 
by Fig.\ref{weights}, in which it is seen that the corrections to the wave functions are
smaller than the zeroth order forms.  

We calculated the ratio of the capture rates and the mixing angle
numerically, and investigated their intensity and photon energy
dependences. We investigated the case of a laser of energy $10^{-3}$ eV and 
intensity $10^{10}$ W/mm$^2$ as the reference values. As shown in Sec.\ref{non_rel_lim}, the
competing ionization processes are negligible for these values. In this situation, the
enhancement mechanism is efficient for $1s$ electrons, which play the main role in capture processes, and the
capture rate becomes larger by a factor of ten. This is due to the facts that the effective mass is an increasing
function of $I/E_\gamma^2$, while the 
tunnel ionization and MPI are suppressed by low energy, low intensity laser light, as discussed in
Sec.\ref{non_rel_lim}. 
One way to produce such low energy, low intensity laser is to boost the atom relatively to the laser
source. When the laser irradiates a moving atom from behind, the energy and intensity of the laser 
experienced by the moving atom are smaller than those that would be experienced by an atom at rest
relative to the source, while $m_{eff}$ remains the same.  
Thus, it is possible to convert a laser with high energy and intensity into one with low energy and
intensity.   

As seen in Fig.\ref{effmass_mixing}, which displays our numerical results, 
on enhancement of the atomic electron capture rate can be of several orders, can be realized even
for an intensity $\sim \mathcal{O}(10^{10})$ W/mm$^2$. As
shown in Fig.\ref{energy_dep}.(a), a low energy laser is suitable
for the enhancement of the capture rate. Figure \ref{energy_dep} (b) shows that
the electron becomes bound more closely to the nucleus and muon as the laser
intensity becomes higher. This can be naively understood from the fact that the
photon number density is higher for lower energy photons as long as the
laser intensity, Eq.\eqref{photon_num}, is fixed. The strength of the 
interaction between an electron
and a photon does not depend on the photon energy. Therefore, a denser
photon bath will result in an electron with a larger effective mass. 
It should be noted that the approximation employed in this paper becomes more
reliable as the energy of the photons decreases, because a smaller energy implies weaker time and space
dependence. 

The most serious question concerning the validity of our calculation regards the relaxation time. We
calculated the static wave function in the photon bath. However, in a 
realistic situation, the laser begins to irradiate the system at some $t=0$  in the 
vacuum. This means that we need to calculate the transition time from the 1s
state in the vacuum to that in this medium. If this transition time is much
longer than $T_{\text{Laser}}$, then our approximation becomes invalid. We can
expect that, with a sufficiently high intensity laser, this time is
sufficiently small, but we need to verify this. The calculation needed to make this verification is
very complicated, and hence we leave it to a future work. 

Electron capture processes are important, because they provide a variety of methods to explore
new physics, like lepton flavor violations and neutrino oscillations. To obtain high statistics,
the only way to enhance the atomic electron capture rate is to increase the probability density of electrons at
the origin. We have shown that laser irradiation makes bound electrons
heavier and hence enhances the
 electron capture rate. Although our analysis assumes a static electromagnetic field and is
based on a perturbative calculation, we believe that enhancement can be realized with this method. 
If the enhancement due to laser irradiation is verified by experiments, it can be directly
applied to search for new physics using capture processes. 

\section*{Acknowledgements}
 The authors would like to thank Professor M.~Yoshimura (Okayama Univ.), T.~Nakajima (Kyoto Univ.) 
 and K.~Nakajima (KEK) for private discussions. 
 The work of J.~S. is partially supported by a Grant-in-Aid for Scientific Research on Priority Area
 No.~18034001 and No.~17740131. 

\appendix
\section{Zeroth-Order Wave Function}\label{0th-wavefunc}
Below we list some of the functions appearing in the main text:
\begin{subequations}
\begin{align}
 R_{nl}(r)&= N_{nl}^R~ e^{-\frac{\rho}{n}}\left(\frac{2}{n}\rho\right)^l
  L_{n-l-1}^{(2l+1)}\big(\frac{2}{n}\rho\big), \\
 N_{nl}^R&=(a_1)^{-\frac{3}{2}}\frac{2}{n^2}\sqrt{\frac{(n-l-1)!}{(n+l)!}},\\
 L_{n-l-1}^{(2l+1)}\big(\frac{2}{n}\rho\big)&=\sum_{r=0}^{n-l-1}(-1)^r
 \comb{n+l}{n-l-1-r}\frac{1}{r!}\left(\frac{2}{n}\rho\right)^r, \quad \text{(Laguerre polynomial)}\\
 \rho&=\frac{r}{a_1},
\end{align}
\end{subequations}
and
\begin{subequations}
\begin{align}
 Y_{lm}(\theta,\phi)&=N_{lm}P^m_l(\cos\theta)e^{im\phi},\quad \text{(spherical harmonics)}\\
 N_{lm}&=\epsilon\sqrt{\frac{2l+1}{4\pi}\frac{(l-|m|)!}{(l+|m|)!}},\\
 \epsilon&=
 \begin{cases}
  (-1)^m, & (m>0) \\
  1, & (m\le 1)
 \end{cases}\\
 P^m_l(u)&=(1-u^2)^{\frac{|m|}{2}}\frac{d^{|m|}}{du^{|m|}}P_l(u),\quad \text{(associate Legendre polynomial)}\\
 P_l(u)&=\frac{1}{2^l l!}\frac{d^l}{du^l}(u^2-1)^l,\quad \text{(Legendre polynomial)},\\
 u&=\cos\theta.
\end{align}
\end{subequations}

\section{Coefficients}\label{formulae}
The following are the explicit forms of the coefficients appearing in the section 3 of the main text:
\begin{align}
 N_{nn'}^{ll'}&=\frac{1}{4}\left(\frac{2}{n}\right)^{l+2} \left(\frac{2}{n'}\right)^{l'+2}
  \sqrt{\frac{(n-l-1)!}{(n+l)!}\frac{(n'-l'-1)!}{(n'+l')!}},\\
 F_{nn'}^{ll'}(p)&=\sum^{n-l-1}_{r=0} \sum^{n'-l'-1}_{r'=0} \comb{n+l}{n-l-1-r}\left(-\frac{2}{n}\right)^r
 \comb{n'+l'}{n'-l'-1-r'} \left(-\frac{2}{n'}\right)^{r'} \nn \\
 &\qquad\qquad\qquad \times\frac{(r+r'+l+l'+p+1)!}{r!r'!} \left(\frac{nn'}{n+n'}\right)^{r+r'+l+l'+p+1}.
\end{align}
\begin{align}
 G_{nn'}^{ll'}(p)&=\int^\infty_0\!d\rho~\rho^{l+l'+p}
 e^{-\frac{n+n'}{nn'}\rho}L_{n-l-1}^{(2l+1)}\big(\frac{2}{n}\rho\big)
 L_{n'-l'-2}^{(2l'+1)}\big(\frac{2}{n'}\rho\big)\nn \\
 &=\sum_{r=0}^{n-l-1} \sum_{r'=0}^{n'-l'-2} \comb{n+l}{n-l-1-r}\left(-\frac{2}{n}\right)^r 
  \comb{n'+l'-1}{n'-l'-2-r'}\left(-\frac{2}{n'}\right)^{r'} \nn \\
 &\qquad\qquad\qquad\times \frac{(r+r'+l+l'+p)!}{r!r'!} 
  \left(\frac{nn'}{n+n'}\right)^{r+r'+l+l'+p+1}.
\end{align}

\end{document}